\documentclass{article}%
\usepackage{amsmath}
\usepackage{amsfonts}
\usepackage{amssymb}
\usepackage{graphicx}%
\providecommand{\U}[1]{\protect\rule{.1in}{.1in}}
\newtheorem{theorem}{Theorem}
\newtheorem{acknowledgement}[theorem]{Acknowledgement}

\begin{document}
\begin{titlepage}
\vspace{.3cm} \vspace{1cm}
\begin{center}
\baselineskip=16pt \centerline{\Large\bf  Mimetic Dark Matter } \vspace{2truecm} \centerline{\large\bf Ali H.
Chamseddine$^{1,2}$\ , \ Viatcheslav Mukhanov$^{3,4,5}$\ \ } \vspace{.5truecm}
\emph{\centerline{$^{1}$Physics Department, American University of Beirut, Lebanon}}
\emph{\centerline{$^{2}$I.H.E.S. F-91440 Bures-sur-Yvette, France}}
\emph{\centerline{$^{3}$Theoretical Physics, Ludwig Maxmillians University,Theresienstr. 37, 80333 Munich, Germany }}
\emph{\centerline{$^{4}$LPT de l'Ecole Normale Superieure, Chaire Blaise Pascal, 24 rue Lhomond, 75231 Paris cedex, France}}
\emph{\centerline{$^{5}$MPI for Physics, Foehringer Ring, 6, 80850, Munich, Germany}}
\end{center}
\vspace{2cm}
\begin{center}
{\bf Abstract}
\end{center}
We reformulate Einstein's theory of gravity, isolating the conformal degree of freedom in a covariant way. This is done
by introducing a physical metric defined in terms of an auxiliary metric and a scalar field appearing
through  its first derivatives. The resulting equations of motion split into a  traceless equation obtained
through variation with respect to the auxiliary metric and an additional differential equation for the trace part.
As a result the conformal degree of freedom becomes dynamical even in the absence of matter. We show that this
extra degree of freedom can mimic cold dark matter.
\end{titlepage}

Consider a physical metric $g_{\mu\nu}$ to be a function of a scalar field
$\phi$ and an auxiliary metric $\tilde{g}_{\mu\nu},$ defined by
\begin{equation}
g_{\mu\nu}=\left(  \tilde{g}^{\alpha\beta}\partial_{\alpha}\phi\partial
_{\beta}\phi\right)  \tilde{g}_{\mu\nu}\equiv P\,\tilde{g}_{\mu\nu}.\label{1}%
\end{equation}
It is obvious that in this case the metric $g_{\mu\nu}$ is invariant with
respect to the conformal transformation of the auxiliary metric $\tilde
{g}_{\mu\nu}$, that is, $g_{\mu\nu}\rightarrow g_{\mu\nu}$ when $\tilde
{g}_{\mu\nu}\rightarrow\Omega^{2}\tilde{g}_{\mu\nu}.$ The action is
constructed in terms of the physical metric $g_{\mu\nu}$, which will be
considered as a function of the scalar field $\phi$ and the auxiliary metric
$\tilde{g}_{\mu\nu},$ that is,
\begin{equation}
S=-\frac{1}{2}\int d^{4}x\,\sqrt{-g\left(  \tilde{g}_{\mu\nu},\phi\right)
}\left[  R\left(  g_{\mu\nu}\left(  \tilde{g}_{\mu\nu},\phi\right)  \ \right)
+\mathcal{L}_{m}\right]  ,\label{3n}%
\end{equation}
where we set $8\pi G=1$ and $\mathcal{L}^{m}$ is the Lagrangian for matter$.$
The action above is obviously invariant under conformal transformation
$\tilde{g}_{\mu\nu}\rightarrow\Omega^{2}\tilde{g}_{\mu\nu}$ because it depends
only on $g_{\mu\nu}$ which is conformally invariant by itself\footnote{We
stress that our considerations are different from other works on tensor-scalar
gravity such as disformal gravity which are extensions of Brans-Dicke type
theories, where the scalar field is dynamical (See e.g. J. D.
Bekenstein,\textit{ }Phys. Rev. \textbf{D48 }(1993) 3641 and for later
developments and references therein M. Zumalacarregui and J. Garcia-Bellido,
arXiv:1308.4685) . In our case, and because of invariance under conformal
transformations of the physical metric, we will show that the scalar field
$\phi$ is equivalent to the scaling factor up to an integrating constant, and
thus is not a new dynamical degree of freedom. }.

Variation of the action is given by
\begin{equation}
\delta S=\int d^{4}x\frac{\delta S}{\delta g_{\alpha\beta}}\delta
g_{\alpha\beta}=-\frac{1}{2}\int d^{4}x\sqrt{-g}\left(  G^{\alpha\beta
}-T^{\alpha\beta}\ \right)  \delta g_{\alpha\beta},\label{4n}%
\end{equation}
where $G^{\mu\nu}=R^{\mu\nu}-\frac{1}{2}Rg^{\mu\nu}$ is the Einstein tensor
and $T^{\mu\nu}$ is the energy momentum tensor for the matter. However, the
variation $\delta g_{\alpha\beta}$ can be expressed in terms of the variation
of the auxiliary metric $\delta\tilde{g}_{\alpha\beta}$ and $\delta\phi,$  and
takes the form%

\begin{align}
\delta g_{\alpha\beta} &  =P\delta\tilde{g}_{\alpha\beta}+\tilde{g}%
_{\alpha\beta}\delta P\nonumber\\
&  =P\delta\tilde{g}_{\alpha\beta}+\tilde{g}_{\alpha\beta}\left(  -\tilde
{g}^{\kappa\mu}\tilde{g}^{\lambda\nu}\delta\tilde{g}_{\mu\nu}\partial_{\kappa
}\phi\partial_{\lambda}\phi+2\tilde{g}^{\kappa\lambda}\partial_{\kappa}%
\delta\phi\partial_{\lambda}\phi\right)  \nonumber\\
&  =P\delta\tilde{g}_{\mu\nu}\left(  \delta_{\alpha}^{\mu}\delta_{\beta}^{\nu
}-g_{\alpha\beta}g^{\kappa\mu}g^{\lambda\nu}\partial_{\kappa}\phi
\partial_{\lambda}\phi\right)  +2g_{\alpha\beta}g^{\kappa\lambda}%
\partial_{\kappa}\delta\phi\partial_{\lambda}\phi,
\end{align}
which implies that
\begin{align}
\delta S &  =-\frac{1}{2}\int d^{4}x\sqrt{-g}\left(  G^{\alpha\beta}%
-T^{\alpha\beta}\ \right)  \nonumber\\
&  \times\left(  P\delta\tilde{g}_{\mu\nu}\left(  \delta_{\alpha}^{\mu}%
\delta_{\beta}^{\nu}-g_{\alpha\beta}g^{\kappa\mu}g^{\lambda\nu}\partial
_{\kappa}\phi\partial_{\lambda}\phi\right)  +2g_{\alpha\beta}g^{\kappa\lambda
}\partial_{\kappa}\delta\phi\partial_{\lambda}\phi\right)  .
\end{align}
The corresponding equations of motion thus become
\begin{equation}
\left(  G^{\mu\nu}-T^{\mu\nu}\ \right)  -\left(  G-T\ \right)  g^{\mu\alpha
}g^{\nu\beta}\partial_{\alpha}\phi\partial_{\beta}\phi=0,\label{einstein}%
\end{equation}%
\begin{equation}
\frac{1}{\sqrt{-g}}\partial_{\kappa}\left(  \sqrt{-g}\left(  G-T\ \right)
g^{\kappa\lambda}\partial_{\lambda}\phi\right)  =\nabla_{\kappa}\left(
\left(  G-T\ \right)  \partial^{\kappa}\phi\right)  =0,\label{phi}%
\end{equation}
where $\nabla_{\kappa}$ denotes the covariant derivative with respect to the
metric $g_{\mu\nu}$. Notice that the auxiliary metric $\tilde{g}_{\mu\nu}$
does not appear in these equations by itself but only via the physical metric
$g_{\mu\nu},$ while the scalar field $\phi$ enters the equations explicitly.
As it follows from (\ref{1}) that
\[
g^{\mu\nu}=\frac{1}{P}\tilde{g}^{\mu\nu},
\]
and therefore the scalar field satisfies the constraint equation
\begin{equation}
g^{\mu\nu}\partial_{\mu}\phi\partial_{\nu}\phi=1.\label{eikonal}%
\end{equation}
Taking the trace of equations (\ref{einstein}) we find that
\begin{equation}
\left(  G-T\ \right)  \left(  1-g^{\mu\nu}\partial_{\mu}\phi\partial_{\nu}%
\phi\right)  =0,
\end{equation}
and this equation is satisfied identically due to (\ref{eikonal}) even for
$G-T\neq0.$ In fact the trace $G-T$ \ is determined by equations (\ref{phi})
and (\ref{eikonal}) and even in the absence of matter, when $T^{\mu\nu}=0,$
the equations for the gravitational field have nontrivial solutions for the
conformal mode. The field $\phi$ satisfies the Hamilton-Jacobi equation for a
unit mass relativistic particle in a gravitational field (\ref{eikonal}) with
the action identified with $\phi$ \cite{LL}. After solving it for $\phi$
equation (\ref{phi}) determines $G-T.$ Thus the gravitational field, in
addition to two transverse degrees of freedom, describing gravitons, acquires
extra longitudinal degree of freedom shared by the scalar field $\phi$ and a
conformal factor of the physical metric. This system however, is constrained
by conformal invariance. To understand what this extra degree of freedom
describes we rewrite equations (\ref{einstein}) in the following form%
\begin{equation}
G^{\mu\nu}=T^{\mu\nu}\ +\tilde{T}^{\mu\nu},
\end{equation}
where%
\begin{equation}
\tilde{T}^{\mu\nu}=\left(  G-T\ \right)  g^{\mu\alpha}g^{\nu\beta}%
\partial_{\alpha}\phi\partial_{\beta}\phi,\label{2}%
\end{equation}
Now compare this expression with the energy momentum tensor for a perfect
fluid%
\begin{equation}
T^{\mu\nu}=\left(  \varepsilon+p\right)  u^{\mu}u^{\nu}-pg^{\mu\nu},\label{3}%
\end{equation}
where $\varepsilon$ is the energy density, $p$ is the pressure and $u^{\mu}$
is four-velocity which satisfies the normalization condition $u^{\mu}u_{\mu
}=1.$ If we set $p=0$ and make the following identification%
\begin{equation}
\varepsilon\equiv G-T,\text{ \ }u^{\mu}\equiv g^{\mu\alpha}\partial_{\alpha
}\phi,\text{\ }\label{4}%
\end{equation}
the energy momentum tensor (\ref{3}) becomes equivalent to $\tilde{T}^{\mu\nu
}.$ Thus, the extra degree of freedom imitate the potential motions of
\textquotedblleft dust\textquotedblright\ with the energy density $G-T$ \ and
the scalar field plays the role of the velocity potential. In the absence of
matter this energy density is equal to $-R,$ which does not vanish for generic
solutions. As one can see the normalization condition for the four-velocity,
$u^{\mu}u_{\mu}=1,$ is equivalent to the scalar field equation (\ref{eikonal})
and the conservation law for $\tilde{T}^{\mu\nu}$ gives%
\begin{equation}
0=\nabla_{\mu}\tilde{T}_{\nu}^{\mu}=\partial_{\nu}\phi\nabla_{\mu}\left(
\left(  G-T\ \right)  \partial^{\mu}\phi\right)  +\left(  G-T\ \right)
\partial^{\mu}\phi\nabla_{\mu}\partial_{\nu}\phi.\label{5}%
\end{equation}
The second term here vanishes because by differentiating $g^{\mu\nu}%
\partial_{\mu}\phi\partial_{\nu}\phi=1$ we get $\partial^{\mu}\phi\nabla_{\nu
}\partial_{\mu}\phi=0$ and $\nabla_{\nu}\partial_{\mu}\phi=\nabla_{\mu
}\partial_{\nu}\phi.$ Therefore the conservation law for $\tilde{T}^{\mu\nu}$
leads to equation (\ref{phi}).

To find the explicit solution of this equation it is convenient to work in
synchronous coordinate system where the metric takes the form%
\begin{equation}
ds^{2}=d\tau^{2}-\gamma_{ij}dx^{i}dx^{j},
\end{equation}
with $\gamma_{ij}$ being a three dimensional metric. Moreover taking the
hypersurfaces of constant time to be the same as the hypersurfaces of constant
$\phi$ (see (\cite{LL}) for details), that is,
\begin{equation}
\phi\left(  x^{\mu}\right)  \equiv\tau,
\end{equation}
we find that (\ref{eikonal}) is satisfied. In turn equation (\ref{phi})
becomes%
\begin{equation}
\partial_{0}\left(  \sqrt{\det\gamma}\left(  G-T\ \right)  \right)  =0,
\label{6}%
\end{equation}
and hence
\begin{equation}
G-T=\frac{C\left(  x^{i}\right)  }{\sqrt{\det\gamma}},
\end{equation}
where $C\left(  x^{i}\right)  $ is constant of integration depending only on
spatial coordinates. In particular in flat Friedman universe, where
\begin{equation}
\gamma_{ij}=a^{2}\left(  \tau\right)  \delta_{ij},
\end{equation}
we have
\[
G-T=\frac{C}{a^{3}},
\]
that is, we have a \textquotedblleft dark matter\textquotedblright\ without
dark matter, which is imitated by extra scalar degree of freedom of the
gravitational field. With respect to the gravitational interaction this new
\textit{mimetic} dark matter behaves precisely in the same way as the usual
dark matter (in particular, it is influenced by the gravitational
instability), but it does not participate in any other interaction besides of
the gravitational one. The \textquotedblleft amount\textquotedblright\ of this
mimetic dark matter is determined by the constant of integration $C\left(
x^{i}\right)  .$

To make the model above realistic in inflationary cosmology it must be
modified. In fact, if inflation lasts longer than 70 e-folds, then for those
initial conditions of $C\left(  x^{i}\right)  $ which do not spoil inflation,
the \textquotedblleft amount\textquotedblright\ of mimetic dark matter
remaining at the end of inflation will be completely negligible. To protect
its energy density from decay during the exponential expansion, $a=H^{-1}%
\exp\left(  Ht\right)  $ one can, for example, introduce the coupling of field
$\phi$ with the inflaton field $\varphi$ of the form
\begin{equation}
\phi F\left(  \varphi\right)  ,
\end{equation}
where $F\left(  \varphi\right)  $ is some function of the inflaton field. In
this case equation (\ref{6}) is modified during inflation as
\begin{equation}
\frac{1}{a^{3}}\partial_{0}\left(  a^{3}\left(  G-T\ \right)  \right)
=F\left(  \varphi\right)  .
\end{equation}
Because the inflaton field is changing slowly the function $F\left(
\varphi\right)  $ can be taken as a constant and the approximate general
solution of this equation is%
\begin{equation}
G-T\approx-\frac{F\left(  \varphi\right)  }{3H}+C\exp\left(  -3Ht\right)
\end{equation}
and at the end of inflation when the second term decays
\begin{equation}
G-T\approx-\frac{F\left(  \varphi\right)  }{3H}%
\end{equation}
It is clear from here that if the inflaton field is slightly inhomogeneous
then the \textit{produced mimetic dark matter} will also be inhomogeneous and
the resulting perturbations will be similar to adiabatic perturbations in case
of real cold dark matter.

\bigskip

\begin{acknowledgement}
We thank Lars Brink and Costas Bachas for useful discussions. The work of AHC
is supported in part by the National Science Foundation Phys-1202671 and by
the Humboldt Foundation. The work of VM is supported by \textquotedblleft
Chaire Internationale de Recherche Blaise Pascal financ\'{e}e par l'Etat et la
R\'{e}gion d'Ile-de-France, g\'{e}r\'{e}e par la Fondation de l'Ecole Normale
Sup\'{e}rieure\textquotedblright, by TRR 33 \textquotedblleft The Dark
Universe\textquotedblright\ and the Cluster of Excellence EXC 153
\textquotedblleft Origin and Structure of the Universe\textquotedblright.
\bigskip
\end{acknowledgement}

\end{document}